\documentclass[preprint]{jpsj2} 

\usepackage{graphicx}
\usepackage{dcolumn}
\usepackage{bm}

\begin{document}

\title{Optical Conductivity and Electronic Structure of 
CeRu$_4$Sb$_{12}$ under High Pressure}   

\author{%
Hidekazu \textsc{Okamura}\thanks{E-mail: okamura@kobe-u.ac.jp}, 
Ryosuke \textsc{Kitamura}, 
Masaharu \textsc{Matsunami}\thanks{Present address: 
UVSOR Facility, Institute for Molecular Science, 
Okazaki 444-8585}, 
Hitoshi \textsc{Sugawara}, 
Hisatomo \textsc{Harima}, 
Hideyuki \textsc{Sato$^1$}, 
Taro \textsc{Moriwaki$^2$}, 
Yuka \textsc{Ikemoto$^2$}, and 
Takao \textsc{Nanba}
}
\inst{%
Department of Physics, Graduate School of Science, Kobe University, 
Kobe 657-8501
\\
$^1$Graduate School of Science, 
Tokyo Metropolitan University, 
Tokyo 192-0397 
\\
$^2$Japan Synchrotron Radiation Research Institute (JASRI) and 
SPring-8, Sayo 679-5198.  
}

\recdate{\today}

\abst{
Optical conductivity [$\sigma(\omega)$] of Ce-filled skutterudite 
CeRu$_4$Sb$_{12}$ has been measured at high pressure to 8~GPa and 
at low temperature, to probe the pressure evolution of its 
electronic structures.     
At ambient pressure, a mid-infrared peak at 0.1~eV was formed in 
$\sigma(\omega)$ at low temperature, and the spectral weight below 
0.1~eV was strongly suppressed, due to a hybridization of the $f$ 
electron and conduction electron states.  
With increasing external pressure, the mid-infrared peak shifts to 
higher energy, and the spectral weight below the peak was further 
depleted.    The obtained spectral data are analyzed in comparison 
with band calculation result and other reported physical properties.   
It is shown that the electronic structure of CeRu$_4$Sb$_{12}$ 
becomes similar to that of a narrow-gap semiconductor under 
external pressure.  
}

\kword{Heavy fermion, filled skutterudite, optical conductivity, 
high pressure}

\maketitle

\section{Introduction}
Physical properties of materials under external pressure 
have attracted much interest recently.\cite{airapt}   
By applying a hydrostatic pressure, one may reduce the 
interatomic distance in a material, and therefore can 
tune its physical properties in a continuous manner.  
In addition, the pressure technique does not introduce 
any disorder into the crystal lattice, unlike the case 
of chemical alloying.   
Various novel properties such as a 
superconductivity \cite{kotegawa}, a transition/crossover 
between localized and delocalized states\cite{ybs}, and 
quantum critical transitions\cite{Ce115} have been explored 
under external pressure.   It is interesting to experimentally 
examine the electronic structures associated with these 
pressure-induced phenomena.   Infrared (IR) spectroscopy 
technique is very useful in this regard, since it can 
give the optical conductivity [$\sigma(\omega)$] of a sample 
loaded in a pressure generating cell.    $\sigma(\omega)$ 
contains much information about the microscopic electronic 
structures near the Fermi level.   Note that the other 
common spectroscopic techniques such as photoemission and 
tunneling spectroscopies are technically difficult to 
perform with a pressure cell.    In this work, we apply 
the high pressure IR technique to CeRu$_4$Sb$_{12}$.

CeRu$_4$Sb$_{12}$ is one of the compounds with the ``filled 
skutterudite'' crystal structure, which have attracted 
a great amount of attention due to their interesting 
physical properties.\cite{sku}   
In the case of Ce-filled skutterudites Ce$M_4X_{12}$, 
the reported physical properties are widely varied 
depending on the constituent atoms $M$ and $X$.   
For $X$=P and As, the compounds studied so far 
($M$=Fe, Ru, and Os) all show semiconductor-like 
properties with the estimated energy gaps ranging 
from 5-10~meV for As compounds\cite{CeRuAs,CeOsAs} 
to 40-130~meV for P compounds.\cite{CeFeP,CeRuP,CeOsP}   
For $X$=Sb, in contrast, physical properties for 
the three compounds with $M$=Fe, Ru, and Os show more 
metallic characteristics.  
 In particular, CeRu$_4$Sb$_{12}$ shows some anomalous 
and interesting 
properties.\cite{CeRuSb,maple,low-P,kurita,kurita2}   
The electrical resistivity ($\rho$) of CeRu$_4$Sb$_{12}$ 
rapidly decreases at low temperature ($T$) below 100~K, 
and its electronic specific heat coefficient is moderately 
enhanced, $\gamma \sim$ 80~mJ/K$^2$mol.  
These properties 
are characteristic of intermediate valence Ce compounds.   
In addition, pronounced non-Fermi liquid properties were 
observed below 4~K.\cite{CeRuSb}   Under external pressure, 
$\rho$ of CeRu$_4$Sb$_{12}$ below 100~K was found to 
markedly increase.   At 8~GPa, the $\rho(T)$ data suggested 
an energy gap of a few meV,\cite{kurita,kurita2} and it 
showed further increases at 10~GPa.\cite{kurita2}    
The optical conductivity $\sigma(\omega)$ of 
CeRu$_4$Sb$_{12}$ in the IR range has also been studied 
at ambient pressure.\cite{dordevic1,matunami1,dordevic2}   
$\sigma(\omega)$ showed a strong suppression below 0.1~eV 
at low $T$,\cite{dordevic1,matunami1} with the tail of a 
narrow Drude response due to heavy electron 
state.\cite{dordevic2}

In this work, in order to probe the pressure evolution 
of electronic structures in CeRu$_4$Sb$_{12}$, we have derived 
its $\sigma(\omega)$ at pressures up to 8~GPa.  
The infrared data are analyzed in comparison with the band 
calculation result and other published physical properties.  
It is shown that the electronic structures of CeRu$_4$Sb$_{12}$ 
indeed becomes similar to that of a semiconductor under 
high pressure.

\section{Experimental}
The samples of CeRu$_4$Sb$_{12}$ used in this work were 
single crystals grown with self-flux method.\cite{low-P}    
$\sigma(\omega)$ spectra in vacuum were obtained from 
the measured optical reflectance spectra [$R(\omega)$].   
$R(\omega)$ in vacuum were measured under near-normal 
incidence configuration between 7~meV and 30~eV, as 
previously discussed in detail.\cite{okamura-YbAl3,matunami2}   
The Kramers-Kronig 
(KK) analysis was used to derive $\sigma(\omega)$ from 
the measured $R(\omega)$ data, where the low-energy 
range below the measurement limit was extrapolated 
with a Hagen-Rubens function.\cite{burns,dressel}    
For high pressure measurements, an external pressure was 
applied with a diamond anvil cell (DAC), as previously 
described.\cite{ybs,okamura-airapt}    In a DAC, a 
mechanically polished surface of a sample with 
approximately 200~$\times$ 200~$\times$ 30~$\mu$m$^3$ 
dimensions was closely attached to the culet face of a 
diamond anvil (See also Fig.~8(a) in the Appendix.)   
The diameter of the culet face was 800~$\mu$m.   The 
diamond anvils were of type IIa with low density of 
impurities, which is crucial for infrared 
studies.\cite{okamura-airapt}   
The pressure transmitting medium used was glycerin, which 
has been shown to have good characteristics as a pressure 
transmitting medium.\cite{tateiwa}    
The $R(\omega)$ of the sample was measured relative to 
a gold film mounted together with the sample in DAC 
[see also Fig.~8(a)].      
To measure $R(\omega)$ of a small sample under such 
a restricted condition, synchrotron radiation (SR) was 
used as a bright IR source at the beam line BL43IR of 
SPring-8.\cite{BL43IR}    
For high pressure data, Drude-Lorentz fitting analyses 
were used to derive $\sigma(\omega)$ from measured 
$R(\omega)$, as discussed later.  
More technical details of the high pressure $R(\omega)$ 
measurement with DAC and IR SR can be found 
elsewhere.\cite{okamura-airapt}

\section{Results and Discussion}
For technical reasons, high pressure measurements of 
$R(\omega)$ with DAC were done with mechanically polished 
surfaces, rather than with as-grown surfaces.   In the 
course of this study, we have realized that $R(\omega)$ 
measured on a polished surface of CeRu$_4$Sb$_{12}$ exhibit 
noticeable differences from previous data measured on 
as-grown surfaces.\cite{dordevic1,matunami1}   Therefore, 
we will first describe $R(\omega)$ and $\sigma(\omega)$ 
data measured on both polished and as-grown surfaces, then 
will discuss the results of high pressure experiment.

\subsection{$R(\omega)$ and $\sigma(\omega)$ measured in 
vacuum with polished and as-grown surfaces}
Figures 1(a) and 1(b) compare the $R(\omega)$ and 
$\sigma(\omega)$ spectra of CeRu$_4$Sb$_{12}$ measured on 
mechanically polished and as-grown surfaces, respectively, 
in vacuum.   
The $\sigma(\omega)$ spectra were derived from the 
measured $R(\omega)$ using the KK analysis.  
%
\begin{figure}[t]
\begin{center}
\includegraphics[width=0.7\textwidth]{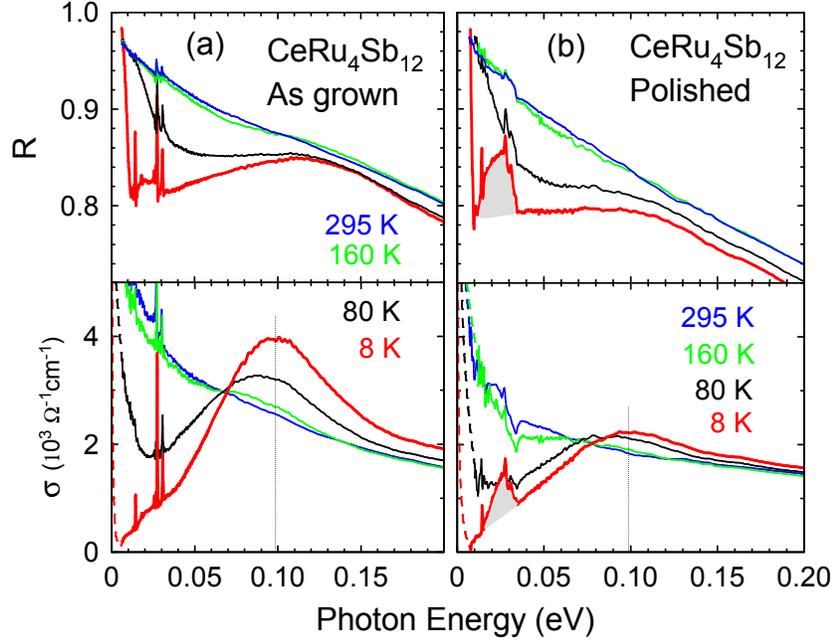}
\end{center}
\caption{(Color online) 
Reflectance (R) and optical conductivity ($\sigma$) of 
CeRu$_4$Sb$_{12}$ in vacuum measured on (a) an as-grown 
surface and (b) a mechanically polished surface.  The 
gray areas in (b) indicate the broadening discussed 
in the text.    The broken curves show the 
extrapolated portion of the spectra, and the vertical 
dotted lines indicate the position of the mid-IR peak 
at 8~K.   The data in (a) are the same as those 
previously reported.\cite{matunami1}
}
\end{figure}
The strong $T$ dependences of $R(\omega)$ and $\sigma(\omega)$ 
have been already discussed in detail for as-grown 
cases.\cite{dordevic1,matunami1}   They result from the 
formation of a conduction (c)-$f$ electron hybridized state.  
Here, important features are the formation of a mid-IR 
peak in $\sigma(\omega)$ at 0.1~eV and the depletion of 
$\sigma(\omega)$ below the mid-IR peak energy.   
The latter feature makes $\sigma(\omega)$ of CeRu$_4$Sb$_{12}$ appear 
like that of an insulator.   However, CeRu$_4$Sb$_{12}$ shows metallic 
characteristics in its transport and magnetic properties at 
low $T$.  There is a narrow, $\delta$-function-like 
Drude component due to heavy-mass carriers below the measurement 
range of this study.\cite{dordevic2}   Such a $\sigma(\omega)$ 
spectrum consisting of a narrow Drude component, a 
depletion of spectral weight (pseudogap), and a mid-IR peak 
have been observed for many intermediate-valence Ce and Yb 
compounds.\cite{matunami2,sievers,hancock,okamura-universal,kimura}  
It has been shown that the mid-IR peak energies observed for 
various Ce compounds are roughly proportional to their 
$c$-$f$ hybridization energies estimated with specific 
heat data and the single-ion 
Anderson model.\cite{dordevic1,hancock,okamura-universal}       
The spectral shapes of our $\sigma(\omega)$ data for the 
as-grown case in Fig.~1(a) agree well with those 
reported by Dordevic {\it et al.}.\cite{dordevic1,dordevic2}

The spectra of the polished surface in Fig.~1(b) are 
basically similar to those of the as-grown one in 
Fig.~1(a), but there are some noticeable differences.  
Most significant is that the optical phonon peaks for 
the as-grown case, seen below 40~meV, are much sharper 
and better resolved than the polished case.     
In the polished case, there is a broad band, indicated 
by the gray area in Fig.~1(b), 
superimposed on the narrower phonon 
lines.  The peak positions of the narrow phonon 
lines, however, have only small differences (less 
than 2~cm$^{-1}$) between the two cases.   
The most likely origin for the broad band is a slight 
disorder in the crystal lattice induced by polishing.  
The disorder may have relaxed the momentum conservation 
rule for a phonon creation by a photon, causing the phonon 
band in Fig.~1(b) due to many phonon modes that are 
forbidden without disorder.  
Nevertheless, the $\sigma(\omega)$ spectra at low 
$T$ of polished case well preserves the main features 
in $\sigma(\omega)$ of the as-grown sample, namely the 
appearance of the mid-IR peak with the spectral depletion 
with cooling.   
Therefore the data obtained from a polished sample should 
be valid for the discussion of electronic structures in 
CeRu$_4$Sb$_{12}$ under pressure.  All the results presented 
hereafter are based on polished samples.

\subsection{$R(\omega)$ measured under high pressure with DAC}   
In a DAC, $R(\omega)$ is measured at a sample/diamond interface, 
rather than the usual sample/vacuum interface.   It should be 
noted that the refractive index of diamond, 2.4, is much larger 
than that of vacuum.    
According to Fresnel's formula, $R(\omega)$ at an interface 
between a material and a transparent medium is expressed 
as:\cite{burns,dressel} 
\begin{equation}
R(\omega)=\frac{(n-n_0)^2+k^2}{(n+n_0)^2+k^2}
\end{equation}
Here, $n$ and $k$ are the real and imaginary parts of the 
complex refractive index of the material, respectively, 
and $n_0$ is the (real) refractive index of the medium, 
$n_0$=2.4 for diamond and 1.0 for vacuum.   
Hereafter, we denote $R(\omega)$ relative to vacuum as 
$R_0(\omega)$, and that relative to diamond as 
$R_{\rm d}(\omega)$.   
Figures~2(a)-2(c) compare measured $R_0(\omega)$, 
$R_{\rm d}(\omega)$ {\it expected} from $R_0(\omega)$, 
and $R_{\rm d}(\omega)$ actually measured in DAC at a 
low pressure of 0.2~GPa, respectively.   
%
\begin{figure}[t]
\begin{center}
\includegraphics[width=0.75\textwidth]{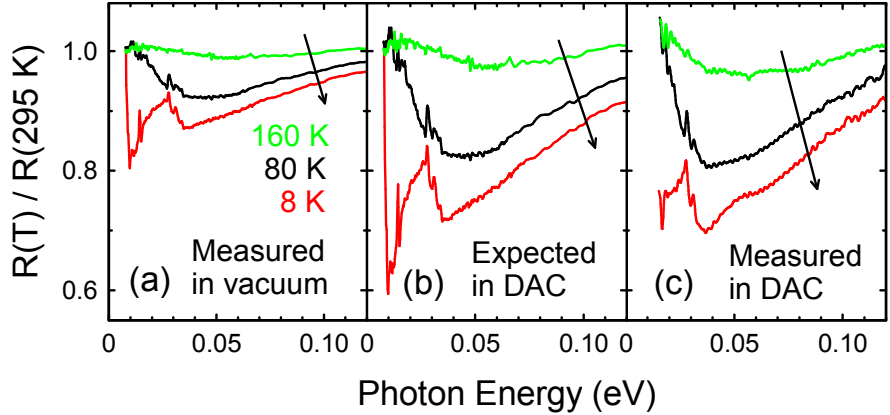}
\end{center}
\caption{(Color online) 
(a) Low energy $R_0(\omega)$ spectra measured 
at low $T$ normalized by that at 295~K.  
(b) $R_{\rm d}(\omega)$ expected from the $R_0(\omega)$ 
spectra in (a), calculated using Eq.~(1), $n_0$=2.4 of 
diamond, and KK analysis as discussed in the text.  
(c) $R_{\rm d}(\omega)$ spectra actually measured in DAC 
at a low pressure of 0.2~GPa.   Note that the low-energy 
limit of measurement is different between (b) and (c).  
}
\end{figure}
To calculate $R_{\rm d}(\omega)$ of Fig.~2(b), $n(\omega)$ 
and $k(\omega)$ were first derived from the KK analysis of 
a $R_0(\omega)$ spectrum, and were substituted into Eq.~(1) 
with $n_0$=2.4 to obtain $R_{\rm d}(\omega)$.    
This was repeated for $R_0(\omega)$ spectra measured at 
different $T$'s.  To highlight their $T$ dependences, 
the spectra in Figs. 2(a)-2(c) have been normalized by 
those at 295~K.   
It is clear that the corresponding spectra in Figs.~2(b) 
and 2(c) agree with each other reasonably well.

Figure~3 shows $R_{\rm d}(\omega)$ spectra measured at 
external pressure of 0, 4, and 8~GPa.   
\begin{figure}[t]
\begin{center}
\includegraphics[width=0.6\textwidth]{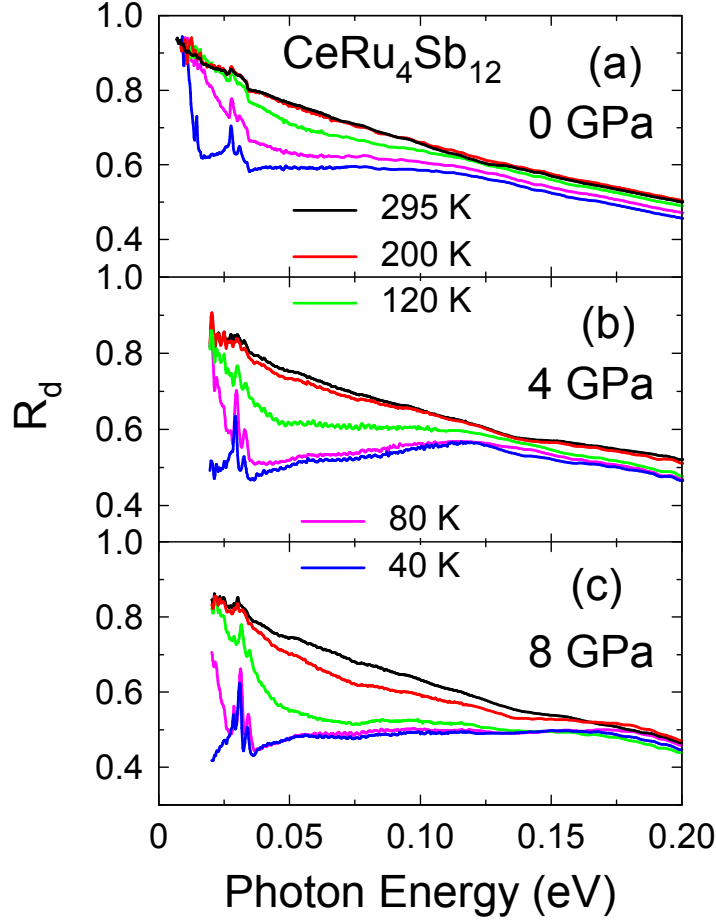}
\end{center}
\caption{(Color online) 
$R_{\rm d}(\omega)$ spectra of CeRu$_4$Sb$_{12}$ 
at various pressures and temperatures.   The spectra at 0~GPa 
in (a) were derived via KK analysis of $R_0(\omega)$ as discussed 
in conjunction with Fig.~2, and those at 4 and 8~GPa in (b) 
and (c) were actually measured in DAC.    
}
\end{figure}
In the present study, $R_{\rm d}(\omega)$ was 
measured with varying $T$ at fixed pressures of 4 and 8~GPa, 
rather than with varying pressure at fixed $T$, since it 
was technically difficult with our gas-driven DAC to 
vary the pressure up to 8~GPa at low $T$.    
$R_{\rm d}(\omega)$ spectra at 0~GPa in Fig.~3(a) are derived 
from $R_0(\omega)$ via KK analysis, as already discussed for 
the spectra in Figs.~2(a) and 2(b).   
The spectra at 4 and 8~GPa in Figs.~3(b) and 3(c) were obtained 
as follows.  
The reflectance spectra above 70~meV were actually measured 
in DAC.  Those below 70~meV, i.e., the far IR range, 
were obtained by multiplying the spectra at 0~GPa in Fig.~3(a) 
by the relative changes of $R_{\rm d}(\omega)$ with 
pressure and temperature measured in DAC.   This procedure for 
the far IR range was taken because it was technically 
difficult to accurately determine the absolute value of 
the reflectance in DAC, due to diffraction effects of long 
wavelength far-IR range.    
The uncertainty in the magnitude of the 
overall $R_{\rm d}(\omega)$ under high pressure is estimated 
to be $\sim$ 2~\%.  
Figure~3 clearly shows that, with increasing pressure, the 
depletion of $R_{\rm d}(\omega)$ with cooling becomes larger, 
and occurs over a wider photon energy.

\subsection{$\sigma(\omega)$ derived from measured 
$R_{\rm d}(\omega)$ at high pressures} 
To obtain the optical conductivity $\sigma(\omega)$ from 
the $R_{\rm d}(\omega)$ spectra measured in DAC, we have 
used Drude-Lorentz (DL) spectral fitting of $R_{\rm d}(\omega)$ 
rather than the KK analysis.   The DL fitting was used 
rather than the KK analysis because the present 
study with DAC is performed only over a limited energy 
range (below 1.1~eV), although a KK analysis generally 
requires a wider spectral range.   
In addition, it is difficult with KK analysis to take 
into account multiple reflections within a thin layer 
of pressure medium, that was present between the sample 
and diamond.   All the details of the DL fitting procedure 
and the analysis of multiple reflections are described 
in the Appendix, with a list of the obtained fitting 
parameters and examples of actual fittings.  Here we 
present the obtained $\sigma(\omega)$ spectra only.

Figure 4 shows the obtained $\sigma(\omega)$ spectra 
of CeRu$_4$Sb$_{12}$.  
\begin{figure}[t]
\begin{center}
\includegraphics[width=0.55\textwidth]{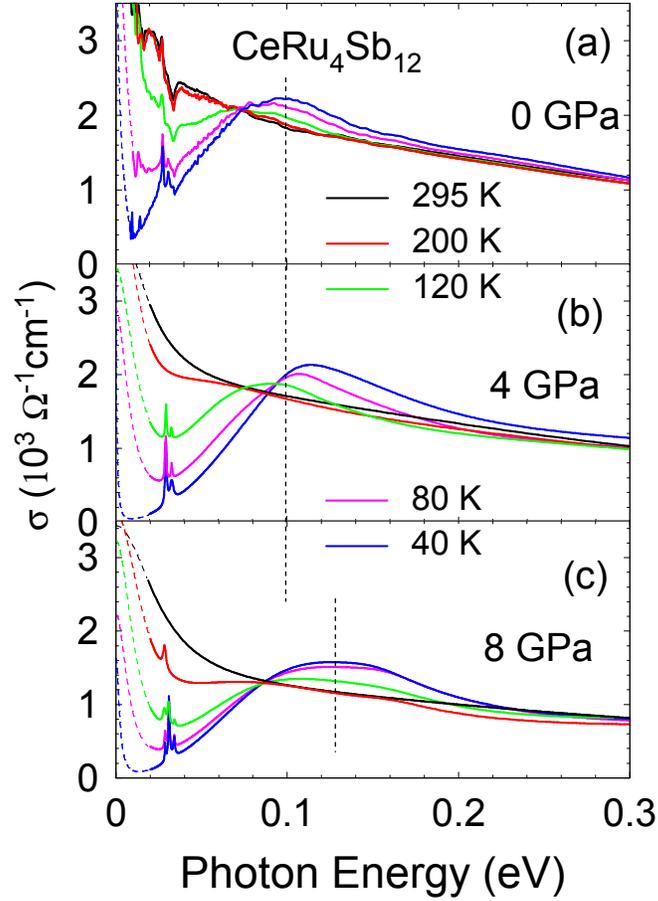}
\end{center}
\caption{
(Color online) 
(a)-(c) show optical conductivity spectra ($\sigma$) 
at external pressures of 0, 4, and 8~GPa, respectively.  
The vertical lines indicate the shift of mid-IR peak 
with pressure.   
The broken curves in (a) indicate 
the extrapolated portion of the spectra, and those 
in (b) and (c) indicate the range below the low-energy 
limit of the measurement.    
}
\end{figure}
The spectra at 0~GPa in (a) were obtained from 
$R_0(\omega)$ with K-K analysis, and those at 4 and 
8~GPa in (b) and (c) were obtained from measured 
$R_{\rm d}(\omega)$ with the DL fitting.   
From the data, the $T$ evolution of $\sigma(\omega)$ 
at high pressure appears qualitatively similar to those 
at zero pressure.   
Namely, a mid-IR peak develops with cooling, and 
the spectral weight at the lower-energy side of the 
peak is progressively depleted with cooling.   
However, under pressure, the 
development of the mid-IR peak starts at higher $T$.   
For example, at 0~GPa there is no mid-IR peak at 200~K 
yet, but at 8~GPa, it is already observed at 200~K.  
In addition, the mid-IR peak apparently shifts to 
higher energy with pressure, 
as indicated by the vertical broken lines in 
Fig.~4, from about 0.1~eV at 0~GPa to 0.13~eV at 8~GPa.  
Figure~5 plots the mid-IR peak energies versus $T$, 
which confirms the above observations.     
\begin{figure}[t]
\begin{center}
\includegraphics[width=0.5\textwidth]{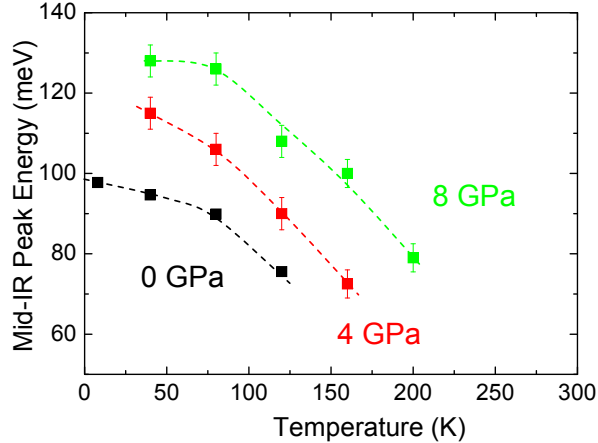}
\end{center}
\caption{(Color online) 
Mid-IR peak position in $\sigma(\omega)$ as a function 
of $T$.   Here, the peak position was determined as the 
approximate center position of the peak, rather than the 
center of a particular oscillator used in the fitting 
(see Appendix).  The error bars indicate estimated 
uncertainty arising from the fitting.   }
\end{figure}
Furthermore, Fig.~4 shows that the spectral depletion 
below the mid-IR peak energy becomes much more pronounced 
at higher pressure.      Combined with our previous 
finding that the mid-IR peak energy is scaled with the 
$c$-$f$ hybridization energy,\cite{okamura-universal} the 
present result indicates that the $c$-$f$ hybridization 
energy actually increases with external 
pressure.    Although this 
property in Ce compounds has been widely assumed on the 
basis of transport and magnetic properties 
measured under external pressure, this is the first, direct 
spectroscopic evidence for the property.

In addition to the mid-IR peak, a shoulder at 25~meV has 
been observed in $\sigma(\omega)$ of 
CeRu$_4$Sb$_{12}$.\cite{dordevic1}   
Namely, $\sigma(\omega)$ below 25~meV decreased rapidly 
with cooling below 80~K, in addition to the overall 
decrease of $\sigma(\omega)$ below the mid-IR peak.   
This shoulder is also seen in Fig.~1(a), although 
it is not clear due to the overlapping phonon peaks.   
Similar shoulder has been observed in $\sigma(\omega)$ 
of many other Ce and Yb 
compounds,\cite{okamura-YbAl3,1-2-10,okamura-YbB12} 
and its relation to $c$-$f$ hybridization state has 
been discussed.   Although the pressure evolution of 
this shoulder would be quite interesting, its energy 
(25~meV) was close to the low-energy limit of our 
study (20~meV), and it was difficult to follow the 
pressure evolution.

\subsection{Analysis of the absorption edge}   
The $\sigma(\omega)$ spectrum at 40~K and 8~GPa shows a 
strong depletion of spectral weight below the mid-IR peak 
energy.  It is indeed very similar to that of a narrow-gap 
semiconductor having a small energy gap and a low density 
of free carriers.   However, unlike the case of an ideal 
band semiconductor,\cite{dressel,cardona} 
$\sigma(\omega)$ in Fig.~4 does not show a clear onset, 
and the energy gap magnitude is unclear.    
The lack of a clear onset may be partly due to the 
nature of Lorentz function itself, since, by 
definition, it goes through the origin.    Therefore, 
to examine the onset of absorption, here we employ a 
technique originally used for indirect-gap 
semiconductors such as Ge,\cite{cardona,Ge} and also 
for Kondo semiconductors.\cite{schle,okamura-YbB12}    
Near the fundamental absorption edge at an indirect 
energy gap, it has been shown that the square root of 
the absorption coefficient is proportional to the 
photon energy relative to energy gap, namely 
$\sqrt{\alpha(\omega)} \propto (\hbar\omega-E_g$).\cite{cardona}  
This dependence is due to indirect (phonon-assisted) 
excitation of electrons by photons.\cite{cardona}   
Figure~6(a) plots $\sqrt{\alpha(\omega)}$ of 
CeRu$_4$Sb$_{12}$ at 40~K and different pressures, and 
Figure~6(b) plots the corresponding $\sigma(\omega)$ 
for comparison.   
\begin{figure}[t]
\begin{center}
\includegraphics[width=0.5\textwidth]{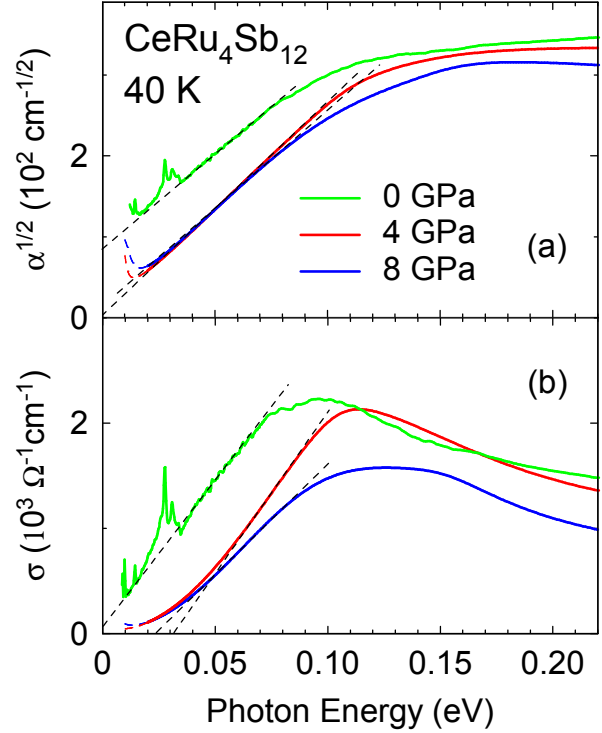}
\end{center}
\caption{(Color online) 
(a) Square root of the absorption coefficient 
($\alpha$) and (b) the optical conductivity 
($\sigma$) of CeRu$_4$Sb$_{12}$ at 40~K and 
external pressures of 0, 4, and 8~GPa.   
The broken lines are guide to the eye 
indicating the linear dependence discussed 
in the text.  
The upturns due to Drude response are not 
shown here for clarity.  
}
\end{figure}
Here, $\alpha(\omega)$ at 0~GPa was obtained from 
the KK analysis of $R_0(\omega)$, while those at 4 
and 8~GPa were derived from the results of DL fitting.  
First, note that at 40~K and 0~GPa, 
$\sqrt{\alpha(\omega)}$ shows a linear energy 
dependence below 80~meV as indicated by the broken 
line.  
As mentioned above, this linear dependence may 
suggest light absorption due to indirect process.  
The intercept of the plot with the energy axis has a 
large negative value, which should correspond to the fact 
that CeRu$_4$Sb$_{12}$ at 0~GPa is a metal, without an energy 
gap.   (Since the original model for indirect 
semiconductors\cite{cardona,Ge} assumed a positive 
intercept with the presence of a gap, the magnitude 
of this negative intercept should not have a physical 
significance.)   
Next, consider the high pressure data at 4 and 8~GPa. 
Since the nature of Lorentz function leads to a linear 
dependence in a $\sqrt{\alpha(\omega)}$ plot, the 
linear dependences alone cannot prove the presence of 
indirect absorption.    However, since such a behavior 
is already observed at 0~GPa without a fitting, it is 
quite reasonable to interpret the data as showing 
pressure-induced shifts of the linear portion already 
observed at 0~GPa.   
Then, note that the intercept of the linear portion 
(dotted line) is near the origin for both 4 and 8~GPa.   
This result may indicate that the overlap between the 
valence band (VB) and conduction band (CB) decreases 
with pressure, and that CeRu$_4$Sb$_{12}$ is close to 
a crossover to a semiconductor.   But since the intercept 
is still near zero energy, there is no clear energy 
gap yet on the basis of the optical spectra.  
This appears consistent with the reported result of 
electrical resistivity.\cite{kurita}   
Namely, a thermally activated 
dependence was observed at 8~GPa, but only with a 
small activation energy ($\sim$ 2~meV) and only 
over a narrow $T$ range (20-40~K).  
Both the very small activation energy and its 
narrow $T$ range indicate that the energy gap 
has not fully opened yet.    
A pressure higher than 10~GPa may be needed 
to open a clear energy gap in CeRu$_4$Sb$_{12}$.  
The $\sigma(\omega)$ spectra in Fig.~6(b) also show a 
nearly linear dependence below the mid-IR peak energy, 
as indicated by the broken lines.   The crossing of the 
linear portion with horizontal axis is located near the 
origin at 0~GPa, but it shifts to about 20~meV at 4 and 
8~GPa.   This result 
also suggests pressure shifts of absorption edge, although 
the linear dependence does not distinguish between indirect 
and direct gaps.

In contrast to the clear shift of mid-IR peak with pressure 
from 4 to 8~GPa in Fig.~4, the absorption edge seen in Fig.~6 
shows almost no shift over the same pressure range.  
This results from a broadening of the mid-IR peak with pressure 
from 4 to 8~GPa.    Although the significance of this result 
is unclear, the broadening probably corresponds to an increase 
of the bandwidth with pressure for the hybridized states 
near $E_{\rm F}$.

\subsection{Electronic structures of CeRu$_4$Sb$_{12}$ 
with and without external pressure}  
At ambient pressure, CeRu$_4$Sb$_{12}$ shows typical 
properties of Ce-based intermediate-valence metals, 
with a rapid decrease of resistivity below 80~K and 
an electronic specific heat coefficient of about 
80~mJ/K$^2$mol.\cite{CeRuSb,maple}   
A Hall effect study showed that the majority carriers 
were the holes,\cite{Hall} and a clear de Haas-van Alphen 
(dHvA) signal showed a Fermi surface with a cyclotron 
mass of about 5~$m_0$.\cite{dHvA}  
These works have suggested that CeRu$_4$Sb$_{12}$ should be a 
semimetal with at least one hole pocket in the Fermi surface.

Figure 7(a) shows the band structure of CeRu$_4$Sb$_{12}$ 
calculated with local density approximation (LDA) and 
full-potential linearized augmented plane wave (FLAPW) 
method.\cite{harima} 
Only the vicinity of Fermi level ($E_{\rm F}$) is shown, 
along the P-$\Gamma$-H line in the Brillouin zone.       
\begin{figure}[t]
\begin{center}
\includegraphics[width=0.75\textwidth]{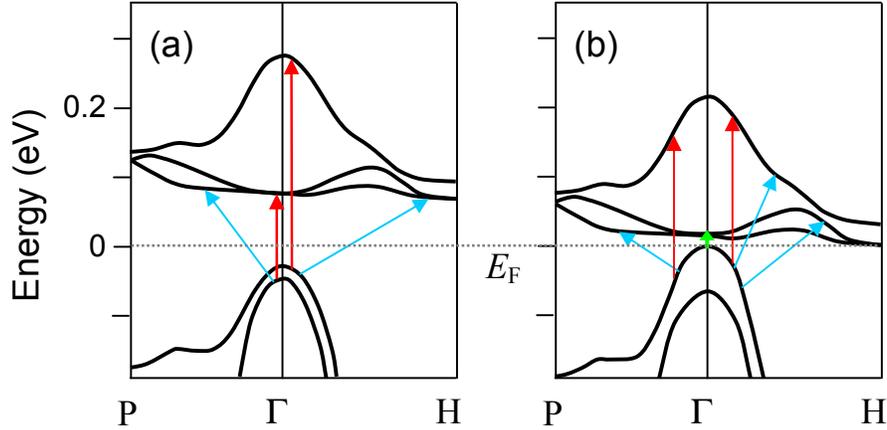}
\end{center}
\caption{
(Color online)  
(a) Band structure of CeRu$_4$Sb$_{12}$ in the 
vicinity of Fermi level ($E_{\rm F}$, dotted line) 
and along the P-$\Gamma$-H line at ambient pressure, 
calculated with LDA and FLAPW.   (b) 
A band structure suggested on the basis of the 
calculated result of (a), the present optical data, 
and the reported metallic characteristics.   
In both (a) and (b), the vertical (red) and 
oblique (blue) arrows indicate 
examples of direct and indirect optical excitations, 
respectively, which may occur in these band structures 
and may contribute to the mid-IR peak in $\sigma(\omega)$.    
The short vertical (green) arrow in (b) shows a ``dark'' 
transition discussed in the text.   
}
\end{figure}
According to the calculation, $E_{\rm F}$ [the broken line in 
Fig.~7(a)] does not cross any band, and hence CeRu$_4$Sb$_{12}$ 
should be an insulator with an energy gap of about 
0.1~eV in the total DOS, between the top of VB at $\Gamma$ 
point and the bottom of CB at H point.    
In the CB, three flat bands are seen around P and H points, 
which originate from the Ce 4$f$ $J$=5/2 levels.  
The dome-shaped portion around $\Gamma$ of the top-most 
band is due to hybridized Sb $p$ state.  The VB is mainly 
derived from the Sb $p$, but the top of VB at $\Gamma$ 
has a strong mixing of Ce 4$f$ state.    
In this situation, the energy gap 
of 0.1~eV in the total DOS should lead to a clear gap in 
$\sigma(\omega)$ with an {\it onset} at 0.1~eV, rather than 
the mid-IR {\it peak} at 0.1~eV as observed.     
[In the case of Ce$X_4$P$_{12}$ ($X$=Fe, Ru, Os),\cite{sku2} 
an energy gap with an onset at 0.2-0.3~eV was actually observed ]  
To explain the above result and the metallic characteristics 
observed in other physical properties, we suggest that 
the actual band structure should be similar to that described 
in Fig.~7(b).   
In this suggested band structure, the flat $f$-derived 
CB states have been moved closer to the VB states.     
Note that, although Fig.~7(b) is drawn as compensated 
semimetal, it is only intended as an illustration, 
assuming a stoichiometric composition.  
In experiments, only one Fermi pocket in the dHvA 
data\cite{dHvA} and a positive Hall coefficient\cite{Hall} 
have been observed.    
These results may indicate a single hole pocket in an 
uncompensated semimetal.    This may possibly result 
from a slight deviation of Ce filling from 100~\%.   
But it has also been discussed that the electron pocket 
at H point, even if present, may be very small and 
may have a very large effective mass, resulting in the 
lack of corresponding dHvA signal and a positive Hall 
coefficient.\cite{Hall,dHvA}

In Figs.~7(a) and 7(b), the vertical (red) and oblique 
(blue) arrows indicate the direct (momentum conserving) 
and indirect (phonon-assisted) optical transitions, 
respectively, which may occur in this band structure 
and may contribute to the mid-IR peak in $\sigma(\omega)$.  
In particular, the indirect ones at low energy range 
suggested by the data in Fig.~6(a) may have actually 
resulted from such transitions as in Fig.~7(b).  
In this case, however, direct transitions from the top of 
VB to the low-lying, $f$-derived CB around $\Gamma$ point, 
shown by the short vertical (green) arrow in Fig.~7(b), 
do not seem to have contributed to the observed $\sigma(\omega)$.  
This is because, if they strongly contribute, $\sigma(\omega)$ 
would have much stronger spectral weight below 0.1~eV.    
One possible reason for such ``dark'' transitions is that 
the states near the top of VB should have strong mixture 
of 4$f$ component, and the transitions such as that 
indicated by the short vertical (green) arrow in Fig.~7(b) 
have a strong $f$-$f$ character:  
The Bloch function of a given state 
should be a superposition of 4$f$-derived component, 
$|f \rangle$, and other components, $| \mbox{else} \rangle$.  
Then the optical transition matrix element should consist 
of the terms $\langle f | \vec{d} | f \rangle $, 
$\langle \mbox{else} | \vec{d} | \mbox{else} \rangle$, 
and $\langle \mbox{else} | \vec{d} | f \rangle$, where 
$\vec{d}$ is a dipole operator.   Since the $f$-$f$ term 
is zero due to the parity selection 
rule,\cite{dressel,cardona} and since the transition shown 
by the short vertical (green) arrow in (b) should have a 
large fraction of such $f$-$f$ transition, the transition 
probability may be also small.  
An effective band model based on a tight binding 
approximation\cite{mutou,saso} was also employed to 
calculate $\sigma(\omega)$ of CeRu$_4$Sb$_{12}$ including 
electron correlation effects.   
By introducing a dispersion (broadening) of 4$f$ electron 
level, a compensated semimetallic band structure was 
obtained.\cite{saso}   This model did not include the 4$f$ 
degeneracy, so that the obtained band structure did not have 
the low-lying, flat CB around $\Gamma$ of Fig.~7(a).    
This model nevertheless reproduced the observed $\sigma(\omega)$ 
qualitatively well, which may also suggest that the 4$f$-derived 
flat bands near $E_{\rm F}$ do not strongly contribute 
to $\sigma(\omega)$.

Under pressure, it is theoretically expected in a Ce compound 
that the $f$ electron levels are shifted upward 
relative to $E_{\rm F}$.    
The CB states shown in Fig.~7 are under strong influence 
of hybridization between Ce 4$f$ and Sb 5$p$ states, and therefore 
it is expected that the CB states are also pushed upward by 
the pressure shift of 4$f$ state.  
This may also reduce the overlap between VB and CB.  
The experimentally observed semiconductor-like characteristics 
of CeRu$_4$Sb$_{12}$ under pressure, namely the increase of 
resistivity, is most likely 
the result of this $f$ electron energy shift.   
This interpretation is also consistent with the results of Fig.~6, 
i.e., an increase in the onset energy of absorption and the 
upward shift of mid-IR peak in $\sigma(\omega)$.   
The upward shift of 4$f$ levels with pressure would also reduce 
the density of thermally populated carriers at low $T$, which 
is consistent with the observed increase of resistivity under 
pressure.\cite{kurita}   
If the pressure is further increased, the 4$f$ level shift would 
eventually lead to the opening of a true energy gap in the total 
DOS, similar to the situation in Fig.~7(a).   
Note that in Ce$M_4$P$_{12}$ and Ce$M_4$As$_{12}$, whose lattice 
constants are smaller than that of Ce$M_4$Sb$_{12}$, an energy 
gap is in fact observed.\cite{CeRuAs,CeOsAs,CeFeP,CeRuP,CeOsP,sku2}

\section{Conclusion}
The optical conductivity $\sigma(\omega)$ of Ce-filled 
skutterudite compound CeRu$_4$Sb$_{12}$ was measured at low 
temperatures under high pressure up to 8~GPa.   
With increasing pressure, the 
characteristic mid-IR peak in $\sigma(\omega)$ shifted toward 
higher energy.  In addition, the depletion of $\sigma(\omega)$ 
below the mid-IR peak energy became more significant and 
well-developed.       
These results were discussed in comparison with other 
experimental results and band calculations.   
The observed evolution of $\sigma(\omega)$ with pressure 
indicates that the energy separation between 
the 4f-derived states above and below the Fermi 
level increases with pressure.  The $\sigma(\omega)$ spectrum 
at 8~GPa is indeed similar to that of a narrow-gap 
semiconductor with a small density of residual 
carriers, which is consistent with the 
semiconductor-like behavior of resistivity with 
a small activation energy over a narrow $T$ range.  
However, on the basis of our optical data, an energy 
gap is not fully open yet at 8~GPa.   A pressure 
greater than 10~GPa may be therefore needed to observe 
a clear gap opening.    Such a higher pressure study 
is desired in future to further probe the electronic 
structures of CeRu$_4$Sb$_{12}$ under pressure.

\section{Acknowledment} 
H. O. would like to thank Dr. H. Yamawaki for 
providing the gasket material used in the DAC.   
This work has been supported by the following 
Grants-In-Aid for Scientific Research from Ministry 
of Culture, Education, Science, Sports and Technology 
of Japan: Innovative Area ``Heavy Electron'' 
(21102512-A01) and Scientific Research B (17340096).  
Experiments at SPring-8 were performed under the 
approval by JASRI (2009B0089, 2009A0089, 2008B1070, 
2008A1239).   

\section{Appendix}  
This Appendix describes the Drude-Lorentz fitting 
procedures used for deriving the optical conductivity 
$\sigma(\omega)$ from the $R_{\rm d}(\omega)$ spectra 
measured in DAC.   Figure~8(a) illustrates the experimental 
condition in the present IR study.    
The fitting procedure takes into account effects of 
diamond refractive index and a thin layer of pressure 
medium which is present between the diamond and 
sample.   These are described in detail below.

\subsection{Drude-Lorentz spectral fitting}
The Lorentz oscillator (classical damped harmonic 
oscillator) model describes the dynamical response 
of bound electrons with a natural (resonance) 
frequency of $\omega_0$ to an electromagnetic wave 
of frequency $\omega$.   According to this model, 
the real ($\epsilon_1$) and imaginary ($\epsilon_2$) 
parts of the complex dielectric function 
are\cite{burns,dressel} 
\begin{equation}
\epsilon_1 = \epsilon_{\infty} + \omega_p^2 
\frac{(\omega_0^2 - \omega^2)}
{(\omega_0^2-\omega^2)^2 + \gamma^2 \omega^2}, 
\end{equation}
\begin{equation}
\epsilon_2 = \omega_p^2 \frac{\gamma \omega}
{(\omega_0^2-\omega^2)^2 + \gamma^2 \omega^2}. 
\end{equation}
Here, $\omega_p$ is the plasma frequency, $\gamma$ is 
the damping (scattering) frequency, and $\epsilon_\infty$ 
is a constant representing the polarizability of 
higher energy electronic transitions.    The Drude model 
for free carriers can be obtained 
by setting $\omega_0$=0.   Complex 
refractive indices $n$ and $k$ can be calculated from 
the relations $\epsilon_1=n^2 - k^2$ and $\epsilon_2=2nk$, 
and $R_{\rm d}(\omega)$ is calculated with Eq.~(1) and 
$n_0$=2.4 for diamond.  
Then $\sigma(\omega)$ is obtained as 
$\sigma(\omega)= \omega \epsilon_2(\omega)/(4\pi)$.  
In the fitting, the parameters $\omega_p$, $\omega_0$ and 
$\gamma$ were varied to minimize 
the difference between the calculated and measured 
$R_{\rm d}(\omega)$ spectra on the basis of 
least-squares fitting scheme.

\subsection{Effects of interference due to 
a thin gap between the sample and diamond}  
In our experimental condition, as shown in Fig.~8(a), 
a very thin layer of pressure medium was present 
between the diamond face and the 
sample surface.  This occurred since the surface of 
the sample was not completely flat while the diamond 
surface was almost perfectly flat.\cite{footnote}    
With such a layer, as shown by red arrows in Fig.~8(a), 
multiple reflections of light and their interference are 
expected.   In fact, the visible image of a sample loaded 
into DAC with pressure medium often showed interference 
fringes (Newton's rings).  
This did not seem to affect the 
measured $R_{\rm d}(\omega)$ in the far IR range, 
as demonstrated by the good agreement between the 
spectra in Fig.~2(b) and 2(c).   In the mid-IR and 
near-IR ranges where the wavelength was shorter, however, 
the ``expected'' 
$R_{\rm d}(\omega)$, calculated from the measured 
$R_0(\omega)$, did show noticeable difference from 
the measured $R_{\rm d}(\omega)$, as shown in Fig.~8(b).   
This shows that the thickness of the medium layer 
was not negligible compared with the wavelength of 
light in the mid-IR range.  
Note that the gold film used as the reference of 
reflectance did not have such a layer, since it was 
directly pressed onto the diamond by the gasket, 
as sketched in Fig.~8(a).     
%
\begin{figure}[t]
\begin{center}
\includegraphics[width=0.6\textwidth]{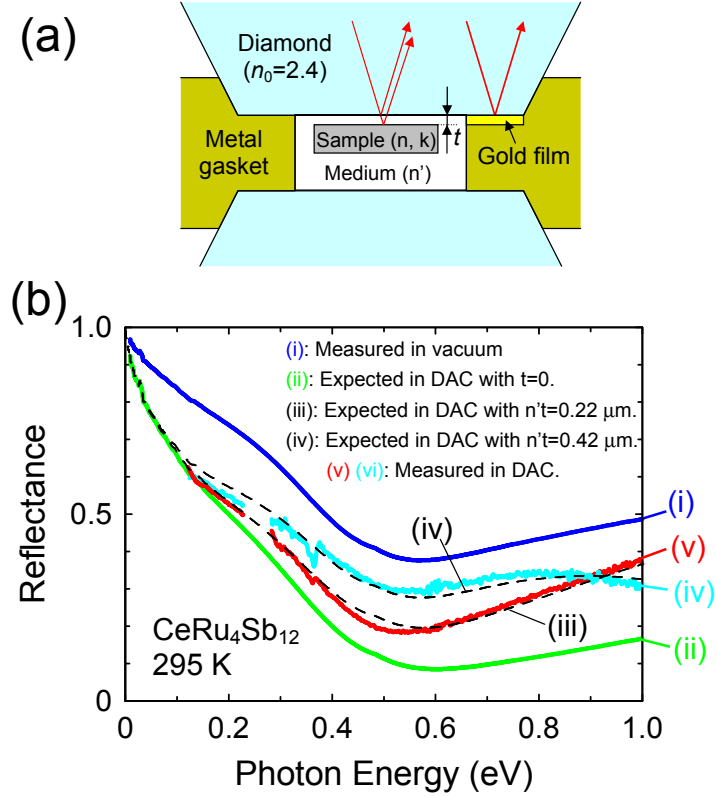}
\end{center}
\caption{(Color online) 
(a) Schematic drawing (not in scale) for the 
experimental condition of our high pressure IR study 
with a DAC.    
 (b) Analysis of interference effects 
on the reflectance spectra.    
(i): $R_0(\omega)$ measured in vacuum at 295~K 
(without using a DAC).  
(ii): $R_{\rm d}(\omega)$ expected from $R_0(\omega)$.  
$R_{\rm d}(\omega)$ was calculated with KK analysis of 
$R_0(\omega)$, Eq.~(1), and $n_0$=2.4 for diamond.  
(iii) and (iv) (broken curves): Spectra calculated 
similarly to (ii), but including the interference 
effects with $n't$=0.22 and 0.42~$\mu$m, respectively. 
(v) and (vi): Two $R_{\rm d}(\omega)$ spectra 
independently measured in the DAC.    
Note that the measured $R_{\rm d}(\omega)$ spectra 
are not shown over a narrow range around 0.25~eV, 
due to strong absorption by diamond.   
}
\end{figure}
%
The reflectance under this situation has been 
theoretically analyzed in detail,\cite{multiple} and 
may be expressed, omitting the $\omega$ dependence, 
as\cite{multiple}    
\begin{equation}
R_d=\left| \frac
{(n_0-n')(n'+ \hat{n}) e^{i\delta} + 
 (n_0+n')(n'- \hat{n}) e^{-i\delta} }
{(n_0+n')(n'+ \hat{n}) e^{i\delta} + 
 (n_0-n')(n'- \hat{n}) e^{-i\delta} }
\right| ^2, 
\end{equation}
where 
\begin{equation}
\delta = \frac{2\pi}{\lambda} n't.  
\end{equation}
Here, $n_0$=2.4 is the refractive index of diamond, 
$n'$ and $t$ are the refractive index and layer 
thickness of the medium, respectively,   
$\hat{n}=n + ik$ is 
the complex refractive index of the sample, and 
$\lambda$ is the wavelength of light in vacuum.   
For simplicity, we assume that $n'$ is a real constant 
(no absorption and dispersion in the medium), and that 
the layer thickness is constant, although in reality 
it may vary over the sample area.  
Then using $n(\omega)$ and $k(\omega)$ of the sample 
obtained 
with the KK analysis of $R_0(\omega)$, we calculated 
$R_{\rm d}(\omega)$ with various values of $n't$.  
In Fig.~9, $R_{\rm d}(\omega)$ spectra calculated 
with $n't$=0.22~$\mu$m (iii) and 0.42~$\mu$m (iv) 
are shown, which agree well with the measured 
spectra (v) and (vi), respectively.   
The magnitude of the obtained $n't$ is quite reasonable since 
the visible image of the sample under microscope exhibited no 
or only one interference fringe when these spectra were measured. 
This means that 2$n't$ should be smaller than or of the same 
order with the visible wavelength, which is consistent 
with the obtained $n't$ values.   
Note also that both the measured and calculated spectra 
indicate that the interference effect becomes 
small below about 0.2~eV.  

For each high pressure run, 
$R_{\rm d}(\omega)$ was measured in DAC before applying 
high pressure, and the obtained value of $n't$ was used 
to analyze the subsequently obtained high pressure data.  
In reality, $n't$ may also change with pressure, but 
it was difficult to correctly measure or estimate it.  	
Therefore we assumed the initial value of $n't$ for 
the analysis of high pressure data.

\subsection{Actual fitting model and examples of fitted spectra}
Once the value of $n't$ was determined, $R_{\rm d}(\omega)$ 
was calculated with a set of parameters, taking into 
account $n_0$=2.4 through Eq.~(1) and the interference effect 
through Eq.~(4).   Then the parameters were adjusted to 
minimize the deviation of the calculated $R_{\rm d}(\omega)$ 
from the measured one.   
The fitting was done with ``RefFIT'', a dedicated 
software for spectral fittings developed by Alexey 
Kuzmenko.\cite{RefFIT}  
Figure~9 shows examples of actual fitting for 
(4~GPa, 40~K) and (8~GPa, 40~K) data.   
\begin{figure}
\begin{center}
\includegraphics[width=0.5\textwidth]{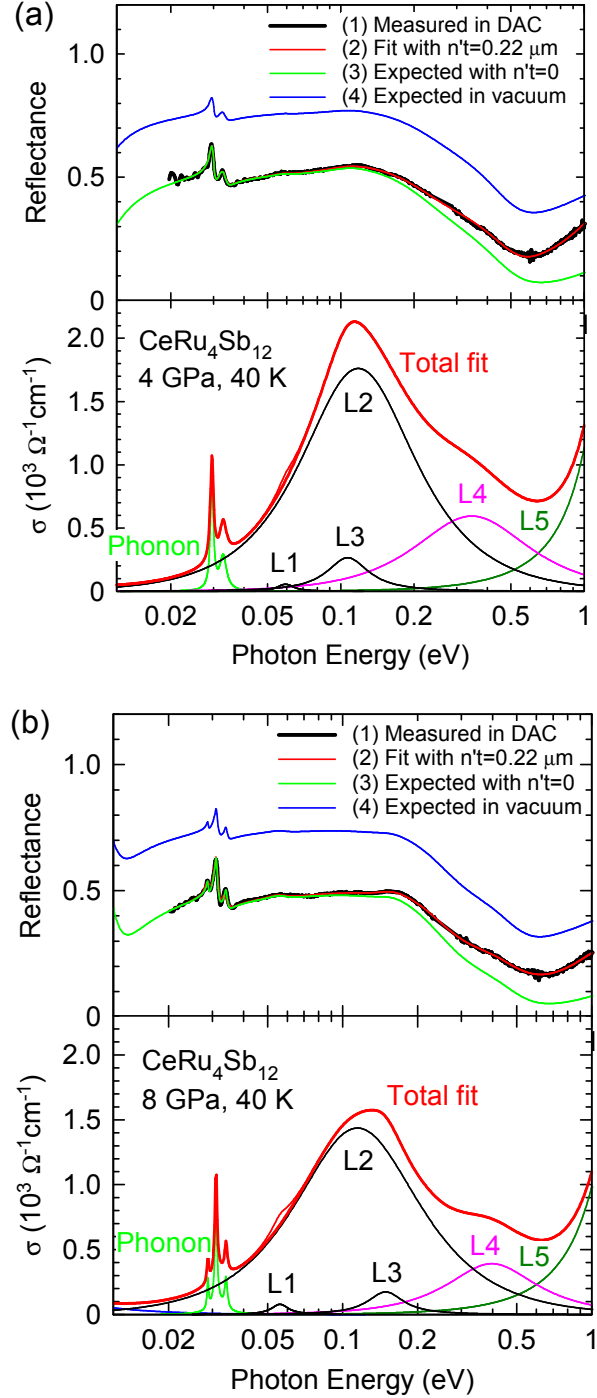}
\end{center}
\caption{(Color online) 
Examples of fitting for $R_{\rm d}(\omega)$ data 
measured in DAC and the resulting optical conductivity 
($\sigma$) at (a) 4~GPa and 40 K and (b) 8~GPa and 40~K.  
(1) Measured $R_{\rm d}(\omega)$ spectrum.  
(2) Fitted spectrum with $n't$=0.22~$\mu$m.    
(3) Spectrum expected in the absence of a medium layer, 
obtained by setting $t$=0 in spectrum (2).  
(4) Spectrum expected in vacuum, obtained by setting 
$n_0$=1 and $t$=0 in spectrum (2).  
The total fit $\sigma(\omega)$ is shown both with (thin red curve) 
and without (thick red curve) the L1 component used to 
cover the hump of $R(\omega)$ at 60~meV, which is an 
instrumental artifact.\cite{footnote3}  
$R_{\rm d}(\omega)$ data between 0.236 
and 0.285~eV, which could not be measured due to strong 
absorption in diamond, was interpolated by a straight 
line.  
}
\end{figure}
Note that the overall spectral shape of 
$R_{\rm d}(\omega)$ is basically reproduced by one 
Drude and three Lorentz (L2, L4 and L5) oscillators.   
Two additional Lorentz oscillators (L1 and L3) were 
used to cover a small hump of $R_{\rm d}(\omega)$ 
near 60~meV and a shoulder of $R_{\rm d}(\omega)$ 
near 0.1~eV (0.17~eV) at 4~GPa (8~GPa), respectively.   
The hump at 60~meV is an instrumental 
artifact,\cite{footnote3} and therefore the $\sigma(\omega)$ 
and $\alpha^{1/2}(\omega)$ spectra shown in Figs.~4 
and 6 were calculated without L1.   
The shoulder became stronger at low $T$, 
and therefore the L3 oscillator was necessary for 
a satisfactory fit.  
The obtained fitting parameters are listed in Table~I.  
For the Drude component, $\omega_p$ and $\gamma$ were 
varied to reproduce $R_{\rm d}(\omega)$, and then 
$\sigma_{\rm opt}$, namely the dc conductivity given by 
these parameters, was calculated and compared with 
the experimentally 
measured $\sigma_{\rm dc}$.\cite{low-P,kurita}   These 
values are also listed in Table~I.    
Note that $\sigma_{\rm opt}$ is higher than $\sigma_{\rm dc}$ 
for most of the data, by a factor of up to 2.1.   This is not 
surprising since a similar result is already seen in the 
data at 295~K and 0~GPa, where the measured 
$\sigma_{\rm dc}$=2600~$\Omega^{-1}$cm$^{-1}$ (Ref.~13) is 
more than twice smaller than that given by the Hagen-Rubens 
extrapolation (Fig.~1).  
Similar result was also observed previously.\cite{dordevic1}   
The reason for this is unclear, but there seems to be some 
scattering mechanism that affects $\sigma_{\rm dc}$ stronger 
than the high frequency one. 
%
\begin{table}
\caption{
Parameters from the Drude-Lorentz fitting of the 
$R_{\rm d}(\omega)$ data, displayed in units of meV 
unless noted.   L1-L5 denote five oscillators as 
indicated in Fig.~9.  $\epsilon_\infty$=4.5 was used 
for all the fitting, and $\omega_0$ of L5 was fixed 
at 1630~meV.   $\sigma_{\rm opt}$ indicates 
the dc conductivity resulting from the fitted Drude 
component, and $\sigma_{\rm dc}$ indicates the actually 
measured values.\cite{low-P,kurita}   
The parameters shown have been rounded to three figures, 
but this does not necessarily mean being accurate to 
three digits.   A blank oscillator means that 
it was not needed to have a good fit.  
}

\includegraphics[width=1.0\textwidth]{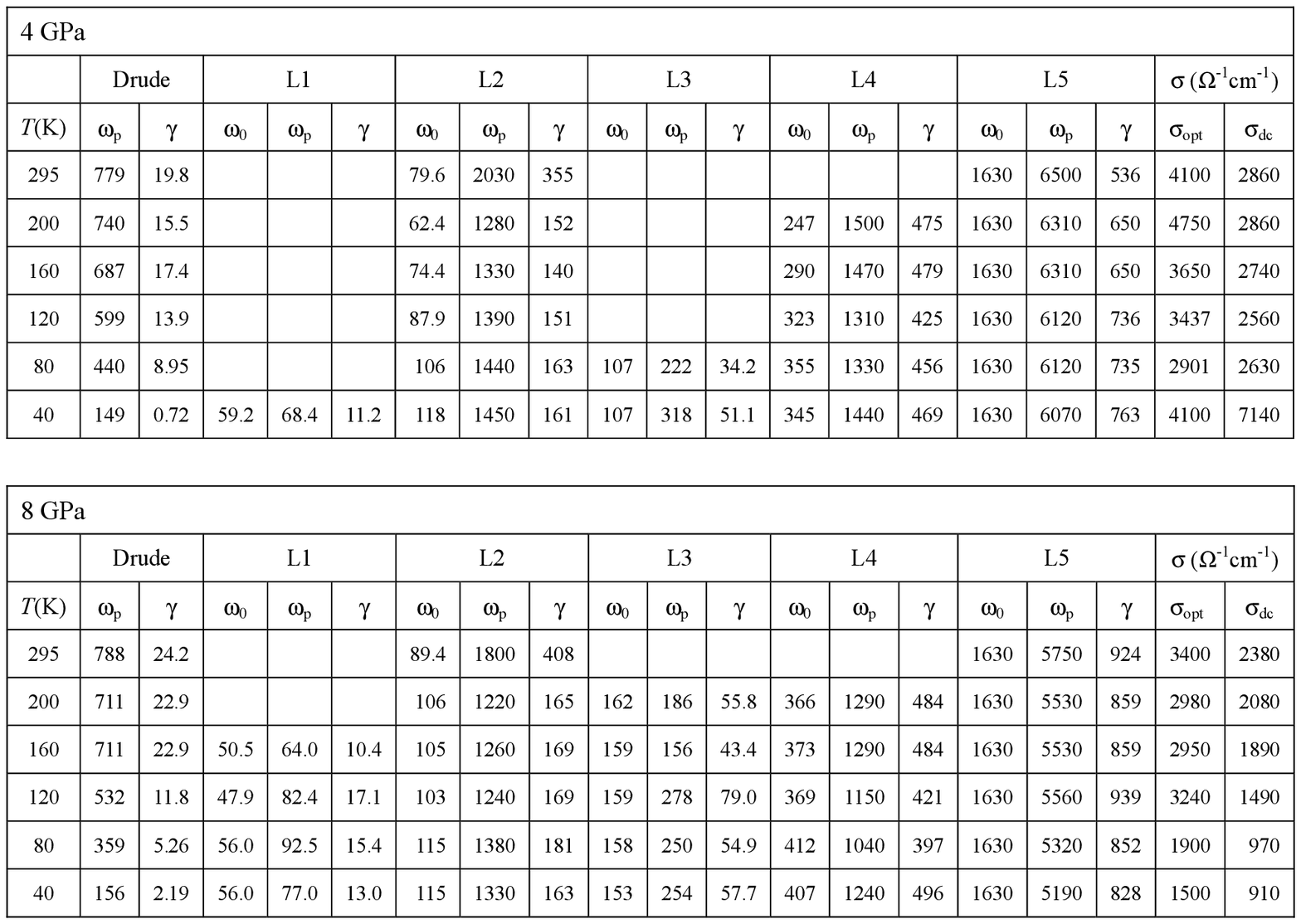}

\end{table}

\subsection{Uncertainty in the magnitude 
of obtained conductivity}
Compared with the case of KK analysis on a reflectance 
measured in vacuum, there are many factors that increase 
the uncertainty in $\sigma(\omega)$ obtained with the 
above procedures.   
The largest source of uncertainty is the arbitrariness 
among the parameters in the DL fitting of $R_{\rm d}(\omega)$ 
performed in a limited energy range (below 1.1~eV).   
The overall uncertainty in the magnitude of resulting 
$\sigma(\omega)$ for the high pressure case is estimated 
to be $\pm$ 10~\%.   
This is indeed much larger than that in the KK-derived 
$\sigma(\omega)$ in vacuum.   However, since we are not 
making any discussion based on the absolute magnitude of 
$\sigma(\omega)$, such as spectral weight transfer or the 
optical sum rule, the uncertainty should not affect 
our conclusions.


\begin{thebibliography}{99}
\bibitem{footnote}   It is possible to avoid such a 
layer of medium between the diamond and sample by 
using a solid pressure medium such as KBr and NaCl.    
At pressures below 10~GPa, however, liquid pressure 
medium such as glycerin can provide much more 
hydrostatic pressure than KBr.\cite{tateiwa}    
In view of the importance of having a more hydrostatic 
pressure for studying electronic structures of strongly 
correlated materials, we chose glycerin over solid medium 
despite the complications in data analysis discussed here.
\bibitem{footnote3} 
This hump of $\sigma(\omega)$ results from a small hump 
of $R_{\rm d}(\omega)$ at 60~meV seen in Figs.~3(b) and 
3(c) at a few temperatures.   To fit this hump, an 
additional oscillator L1 was used as shown in Fig.~9 
(Appendix).   
The hump in $R(\omega)$ is also present in the low 
pressure (0.2~GPa) DAC data of Fig.~2(c), but 
not in the zero pressure data of Figs~1, 
2(a) and 2(b).   We conclude that this hump is 
an instrumental artifact created by the spectroscopy 
apparatus used for the high pressure experiment, 
presumably due to a structure in the spectral 
response of the beam splitter in the IR 
interferometer.
\end{thebibliography}
\end{document}